\documentclass{amsproc}

\newtheorem{theorem}{Theorem}[section]

\theoremstyle{definition}

\theoremstyle{remark}

\numberwithin{equation}{section}

\newcommand{\rz}{\mathbb{R}}       

\begin{document}

\title[binding threshold for the Pauli--Fierz operator]
{Binding threshold for the Pauli--Fierz operator}
\author{Rafael D. Benguria}
\address{Department of Physics, P. Universidad Cat\'olica de Chile, 
Casilla 306, Santiago 22, Chile}
\email{rbenguri@fis.puc.cl}
\thanks{This work  was supported in part by Fondecyt (Chile), 
Project \# 102--0844}
\author{Semjon A. Vugalter}
\address{Matematisches Institut, LMU M\"unchen, Theresienstrasse 39,
80333 Munich, Germany}
\email{wugalter@mathematik.uni-muenchen.de}
\thanks{Work partially supported by HPRN--CT--2002--0027, and the
Volkswagen Stiftung through a cooperation grant}

\date{June 14, 2004}


\begin{abstract}
For the Pauli--Fierz operator with a short range potential we 
study the binding threshold $\lambda_1(\alpha)$ as a function
of the fine structure constant $\alpha$ and show that 
it converges to the binding threshold for the Schr\"odinger
operator in the limit $\alpha \to 0$. 
\end{abstract}

\maketitle

\section{Introduction}

Recently it was proved that the interaction of matter with 
a quantized radiation field leads to the so--called 
{\it enhanced binding} (see \cite{ChVuVu03}; other results with some 
additional restrictions can be found in \cite{HaVuVu03} and \cite{CaHa04}). 
This means that the Pauli--Fierz operator,
which describes a particle interacting with an external potential and
a quantized radiation field may have a ground state even if the
corresponding Schr\"odinger operator (with the same potential)
does not have discrete eigenvalues.

On physical grounds one expects the Schr\"odinger operator to give 
a relatively accurate description of the process of binding of  a 
particle by an external potential, which requires only small
corrections.
This intuition is based on the 
fact that the coupling of the particle with a quantized radiation field 
includes a small parameter (the {\it fine structure constant} $\alpha$), 
and consequently the effects of enhanced binding cannot be very strong.
However a mathematically consistent proof of the fact that for
small $\alpha$ the binding threshold for the Pauli--Fierz operator
is close to the one for the Schr\"odinger operator, as well as the fact of 
the existence of this thresholdup, to the extent of our knowledge, has not been given yet.  
The goal of this paper is to provide such a proof. 

Let
\begin{equation}
H=T + \lambda V,
\label{eq:A}
\end{equation}
be the Pauli--Fierz operator, with the self--energy  operator $T$, 
negative short--range potential $V$ and coupling constant $\lambda$.
And let,
\begin{equation}
h=-\Delta + \lambda V,
\label{eq:B}
\end{equation}
the corresponding Schr\"odinger operator. 
Let $\lambda_0$ be the critical coupling constant such that for
$\lambda > \lambda_0$ the operator $h$ has a ground state and for
$0< \lambda < \lambda_0$ doesn't. Similarly, let $\lambda_1(\alpha)$ 
be the critical coupling constant for the operator $H$. We prove 
that 
$$
\lim_{\alpha \to 0} \lambda_1(\alpha) = \lambda_0.
$$
Given the binding condition 
by Griesemer, Lieb and Loss \cite{GrLiLo01}, the proof of 
{\it enhanced binding} for the Pauli--Fierz operator,
for small $\alpha$, relies
on the construction of a trial function for which the
expectation value of this operator is less than the self--energy
of the particle. This construction proves that $\lambda_1(\alpha)
< \lambda_0$, but it cannot answer the question of how close
is $\lambda_1(\alpha)$ to $\lambda_0$. To answer one needs 
to estimate lower bounds on the quadratic form of $H$. 
In the work at hand this is done by carefully studying 
the properties of the self--energy operator for different values 
of the total momentum. It turns out that very large values of the
total momentum as well as values of the total momentum much smaller
than $\alpha$ are almost irrelevant. For intermediate values of the
momentum we estimate the expectation value of the Pauli--Fierz 
operator in terms of the Schr\"odinger operator with an effective
potential which approaches the original potential when $\alpha$ 
tends to  zero.

\section{Statement of the problem}

The Hamiltonian for an electron interacting with the quantized radiation
field and a given external potential $V(x)$, $x \in \rz^3$, is
\begin{equation}
H=T+ \lambda V(x),
\label{eq:1}
\end{equation}
where
\begin{equation}
T=(p+\sqrt{\alpha} A(x))^2 + g \sqrt{\alpha} \sigma \cdot B(x) + H_f.
\label{eq:2}
\end{equation}
We fix units such that $\hbar=c=1$ and the electron mass $m=1/2$, 
$\alpha=e^2$ is the {\it fine structure} constant, where $e$ is the
charge of the electron. The natural value of $\alpha$ is approximately
$1/137$, however, as usual, we will think about $\alpha$ as a parameter
in the operator $T$. The parameter $g$, whose value is either $0$ or $1$ 
is introduced to describe both the spin ($g=1$) and the spinles ($g=0$)
cases. As usual $\sigma=(\sigma_1, \sigma_2, \sigma_3)$ is the vector
of Pauli matrices, $p= -i \nabla_x$, $B(x)=\nabla \times A(x)$. The magnetic 
vector potential $A(x)$ is given by
\begin{equation}
A(x) = \sum_{\lambda=1,2} \int_{\rz^3}
\frac{\tilde \chi(|k|)}{2 \pi |k|^{1/2}} \varepsilon_{\lambda}
\left[a_{\lambda}(k) e^{i k x}+ a_{\lambda}^*(k) e^{-ikx} \right] \, dk,
\label{eq:3}
\end{equation}
where the operators $a_{\lambda}$, $a_{\lambda}^*$ satisfy the usual
commutation relations
$$
[a_{\nu}(k), a_{\lambda}^*(q)]=\delta(k-q) \delta_{\lambda,\nu}, 
\qquad [a_{\nu}(k), a_{\lambda}(q)]=0.
$$
The vectors $\varepsilon_{\lambda}(k) \in \rz^3$ are the two possible 
orthonormal polarization vectors, perpendicular to $k$. 

The function $\tilde \chi(\vert k \vert)$ in (\ref{eq:3}) describes the 
ultraviolet cutoff on $k$. We assume that $\tilde \chi$ is differentiable and
$\tilde \chi(\vert k \vert) = 0$ for $\vert k \vert > \Lambda$ with
some $\Lambda >0$.

The photon field energy $H_f$ is given by
$$
H_f = \sum_{\lambda=1,2} \int_{\rz^3} |k| a_{\lambda}^*(k) a_{\lambda}(k)\, dk.
$$
We denote by $E_0$ the infimum of the spectrum of $T$.

For $g=1$ the operators $T$ and $H$ are considered on the space
$$
\mathcal{H} = \mathcal{L}_2(\rz^3;C^2) \otimes \mathcal{F},
$$
where $\mathcal{F}$ is the Fock space for the photon field.

If $g=0$, the corresponding space is
$$
\mathcal{H} = \mathcal{L}_2(\rz^3) \otimes \mathcal{F}.
$$
The operator $H$ is semibounded from below and essentially self--adjoint
\cite{BaFrSi99} (see also \cite{Hi02}).

The potential $\lambda V(x)$ is piecewise  
continuous and short--range in the sense
that
\begin{equation}
|V(x)| < c_L(1+|x|)^{-2-\delta}
\label{eq:12}
\end{equation}
for some $\delta >0$ and $c_L>0$, and $\lambda$ is a coupling constant.

The main result of this article will also hold for potentials having
local singularities. However, for the sake of simplicity we will
restrict to piecewise continuous potentials satisfying 
(\ref{eq:12}) in the sequel. 

Together with the operator $H$ we will consider a Schr\"odinger
operator
\begin{equation}
h=-\Delta + \lambda V(x),
\label{eq:13}
\end{equation}
acting on $L^2(\rz^3)$, with the same potential $\lambda V(x)$ as above.

We will assume that $V_{-}(x) = \max(-V(x),0)$ is not identically zero. 
In this case, there is a {\it critical coupling constant} $\lambda_0$
such that for $\lambda \in [0, \lambda_0]$ the discrete spectrum of $h$ is
empty and for $\lambda > \lambda_0$, the discrete spectrum of $h$
is not empty.

Similarly, let $\lambda_1>0$ be the infimum value of $\lambda$ such that
dor all $\lambda > \lambda_1$, $H$ has a ground state. As it was proved
recently (\cite{ChVuVu03}, \cite{HaVuVu03}, \cite{CaHa04})
$\lambda_1< \lambda_0$. Obviously $\lambda_1$ is a function of the fine
structure constant $\alpha$. The main result of this paper is the 
following theorem

\begin{theorem}
Under the conditions stated above,
\begin{equation}
\lim_{\alpha \to 0} \lambda_1(\alpha) = \lambda_0.
\label{eq:14}
\end{equation}
\end{theorem}

\section{Proof of the Theorem}

\subsection{Preliminaries}

According to \cite{ChVuVu03}, \cite{HaVuVu03}, \cite{CaHa04},
 for $\alpha$ sufficiently small
$\lambda_1(\alpha) < \lambda_0$. 
Assume  that 
$\lim_{\alpha \to 0} \lambda_1(\alpha) = \lambda_0$
does not hold. Then, there
exist a constant $\gamma >0$, independent of $\alpha$, and
a sequence $\alpha_n \to 0$, $\alpha_n \in (0,\alpha_0]$, 
for some $\alpha_0>0$,
such that
$H_{\gamma} = T+(\lambda_0 - \gamma) V$ has a groundstate for
all $\alpha_n$. Let 
$\psi_0(\alpha)$ be this groundstate. Then, $(V \psi_0,\psi_0)<0$,
for if $(V \psi_0,\psi_0)=0$ then $\psi_0$ would be an eigenfunction 
of $T$, which is impossible, and on the other hand if $(V \psi_0, \psi_0)>0$
then $\inf \sigma(H_{\gamma}) > \inf \sigma(T)$ which is also impossible.
Consequently,
\begin{equation}
\inf \sigma(H_{\gamma/2}) \le  (H_{\gamma/2} \psi_0, \psi_0) =
(H_{\gamma} \psi_0, \psi_0) + \frac{\gamma}{2}(V \psi_0, \psi_0) < E_0,
\label{eq:15}
\end{equation}
for all $\alpha_n$. 

Now we shall show that (\ref{eq:15}) does not hold for small $\alpha$.
Following (\cite{Fr74}) we introduce the total momentum $P$
\begin{equation}
P=p_{el} \otimes I_f + I_{el} \otimes P_f,
\label{eq:16}
\end{equation}
where $p_{el}$ and $P_f = \sum_{\lambda=1,2} \int_{\rz^3} k \, 
a_{\lambda}^*(k) a_{\lambda}(k) \, dk$ denote the electron and the
photon momentum operators, respectively. The operator $T$ is translationally
invariant, and therefore it commutes with $P$. We will now make a 
partition of unity on the space of total momentum. 
Let $\chi_1=\chi(\vert P \vert < \alpha^q)$, $\chi_3=\chi(|P|>P_0)$,
and $\chi_2=1-\chi_1-\chi_3$, where $P_0>0$, and $q>0$ will be chosen later. 
Here $\chi(S)$ is the characteristic function of the set $S$. 

For an arbitrary state $\psi$, using Schwarz's inequality we get
\begin{equation}
\left(H_{\gamma/2} \psi,\psi \right) \ge \sum_{i=1}^3 L_i [ \psi_i ],
\label{eq:17}
\end{equation}
where $\psi_i = \psi \chi_i$, and
\begin{equation}
L_i[\psi_i]=(T \psi_i, \psi_i) + (\lambda_0-\frac{\gamma}{2}) 
(V \psi_i, \psi_i) -\frac{4}{\kappa}(|V| \psi_i, \psi_i),
\label{eq:18}
\end{equation}
for $i=1,3$, and 
\begin{equation}
L_2[\psi_2]= (T \psi_2, \psi_2) + (\lambda_0-\frac{\gamma}{2}) 
(V \psi_2, \psi_2) -\kappa(|V| \psi_2, \psi_2),
\label{eq:19}
\end{equation}
with $\kappa>0$.

We shall prove 
\begin{equation}
L_i [ \psi_i ] \ge E_0 \vert| \psi_i \vert|^2,
\label{eq:20}
\end{equation}
for $i=1,2,3$ and $\alpha$ sufficiently small. 

In the sequel we choose $\kappa$ such that
\begin{equation}
\sigma_d\left( -\frac{\gamma}{4 \lambda_0} \Delta - \kappa c_L
(1+|x|)^{-2-\delta} \right)= \emptyset,
\label{eq:21}
\end{equation}
which is always possible since $(1+|x|)^{-2-\delta}$ is short--range. 

\subsection{Large momentum estimates}

First notice that using (\ref{eq:12}) we have the estimate
\begin{equation}
L_3 [\psi_3] \ge (T\psi_3,\psi_3) - \tilde c((1+\vert x \vert)^{-2-\delta}
\psi_3, \psi_3),
\label{eq:a1}
\end{equation}
where $\tilde c = (\lambda_0 - (\gamma/2) + (4/\kappa)) c_L$. 

Using the definition of $T$ we have 
\begin{eqnarray}
& (T \psi_3,\psi_3) = - (\Delta_x \psi_3, \psi_3) + 2 
\sqrt{\alpha} \Re (\nabla_x \psi_3, A \psi_3) \nonumber \\
&+\alpha(A^2 \psi_3,\psi_3) + \sqrt{\alpha} g \sigma \cdot (B \psi_3,\psi_3)
+(H_f \psi_3,\psi_3).
\end{eqnarray}
Proceeding in a similar way as in the subsection ``a priori estimates''
on reference \cite{HaVuVu03} (see also \cite{GrLiLo01}, page 586) we get, 
\begin{equation}
|\sqrt{\alpha}(\nabla_x \psi_3, A \psi_3)|
\le c \sqrt{\alpha} \vert| \nabla_x \psi_3 \vert|^2 + c \sqrt{\alpha}
(H_f \psi_3, \psi_3)
\label{eq:a3.1}
\end{equation}
and
\begin{equation}
|\sqrt{\alpha}(B \psi_3, \psi_3)|
\le c \sqrt{\alpha} \vert| \psi_3 \vert|^2 + c \sqrt{\alpha}
(H_f \psi_3, \psi_3).
\label{eq:a3.2}
\end{equation}
Hence, for small $\alpha$ we have
\begin{equation}
(T \psi_3, \psi_3)  \ge (1-c \sqrt{\alpha}) \vert| \nabla_x \psi_3 \vert|^2
+(1-c\sqrt{\alpha}) (H_f \psi_3, \psi_3) - c \sqrt{\alpha} 
\vert| \psi_3 \vert|^2.
\label{eq:a4}
\end{equation}
Here, and in the sequel, $c$ denotes a generic positive constant.
Notice that $\vert| \nabla_x \psi_3 \vert|^2 = \vert| (P-P_f) \psi_3 \vert|^2$,
and $(H_f \psi_3, \psi_3) \ge (|P_f| \psi_3, \psi_3)$. On the support of
the function $\chi_3$, $|P|>P_0$, which implies
$$
(P-P_f)^2 + |P_f| \ge \frac{1}{4} P_0,
$$
assuming $P_0$ is chosen to be greater than $1$. In fact, if $|P_f| 
\ge P_0/2$, then $(P-P_f)^2+|P_f| \ge P_0/2 > P_0/4$. On the other hand,
if $|P_f| \le P_0/2$, then 
$(P-P_f)^2 + |P_f| \ge P_0^2/4 \ge P_0/4$ (since $P_0>1$). 

Therefore,
\begin{equation}
(T \psi_3,\psi_3) \ge (1-\sqrt{\alpha} C) \frac{P_0}{4} 
\vert| \psi_3 \vert|^2 - c \sqrt{\alpha} \vert| \psi_3 \vert|^2. 
\label{eq:a5}
\end{equation}
Using the bound (\ref{eq:a5}) in (\ref{eq:a1}) we finally get
\begin{eqnarray}
& L_3[\psi_3] 
\ge (1-\sqrt{\alpha} C) \frac{P_0}{4} 
\vert| \psi_3 \vert|^2 - c \sqrt{\alpha} \vert| \psi_3 \vert|^2
-\tilde c ((1+\vert x \vert)^{-2 - \delta} \psi_3, \psi_3) \nonumber \\
& \ge  \vert| \psi_3 \vert|^2 \left((1-\sqrt{\alpha} C) \frac{P_0}{4} 
- c \sqrt{\alpha} - \tilde c \right) > E_0(\alpha) \vert| \psi_3 \vert|^2, 
\label{eq:a6 }
\end{eqnarray}
for sufficiently large $P_0$.

\subsection{Small momentum estimates}

To bound $L_1[\psi_1]$ we use the following estimate (see, e.g., 
\cite{ChVuVu03,Ch01}) for sufficiently small $\vert P \vert$ and
all $\alpha \in (0,\alpha_0]$
\begin{equation}
(T \psi_1, \psi_1) \ge E_0 \vert| \psi_1 \vert|^2 + 
\left(1 - d(\alpha)\right) (|P|^2 \psi_1,\psi_1),
\label{eq:b1}
\end{equation}
where $\lim_{\alpha \to 0} d(\alpha)=0$. We will assume $(1-d(\alpha))>3/4$.
The bound (\ref{eq:b1}) in turn implies,
\begin{equation}
L_1[\psi_1] \ge 
E_0 {\vert| \psi_1 \vert|}^2 + 
\frac{3}{4} 
(|P|^2 \psi_1,\psi_1) - 
\tilde c \left((1+\vert x \vert)^{-2-\delta} \psi_1,\psi_1 \right).
\label{eq:b2}
\end{equation}
Let $\tilde \psi_1$ be the function $\psi_1$ written in the 
relative coordinates (see, \cite{ChVuVu03}). 
Namely, as an element of $\mathcal{H}$ the function
$$
\psi_1 = \oplus_{n=0}^{\infty} \psi_n^1(x,s,y_1,\dots, y_n,\lambda_1,
\dots, \lambda_n),
$$
where $s$ is the spin of the particle, $x$ its position vector,
$y_1, \dots, y_n$, the position vectors of photons and $\lambda_1, \dots,
\lambda_n$ the corresponding polarizations. Instead of the variables
$y_1, \dots, y_n$, we introduce the variables $\eta_i=y_i-x$, 
$i=1, \dots, n$ and express the function $\psi_1$ in the new variables
setting 
$\tilde \psi_n^1(x,s,\eta_1,\dots,\eta_n,\lambda_1,\dots,\lambda_n)  
\equiv 
\psi_n^1(x,s,y_1,\dots, y_n,\lambda_1,
\dots, \lambda_n)$ and
$\tilde \psi_1 =\oplus_{n=0}^{\infty} 
\psi_n^1(x,s,y_1,\dots, y_n,\lambda_1,
\dots, \lambda_n)$.

In these new variables
$
(|P|^2 \psi_1, \psi_1) = \vert| \nabla_x \tilde \psi_1 \vert|^2,
$
and 
\begin{equation}
L_1[\psi_1] \ge 
E_0 {\vert| \psi_1 \vert|}^2 + 
\frac{3}{4} \vert| \nabla_x \tilde \psi_1 \vert|^2 - 
\tilde c \left((1+\vert x \vert)^{-2-\delta} {\tilde \psi_1},
{\tilde \psi_1} \right),
\label{eq:b3}
\end{equation}
and $\tilde \psi_1$ is supported in the region $|P| < \alpha^q$.

Define the operator,
\begin{equation}
M \equiv \tilde c |p|^{-1} (1+ |x|)^{-2-\delta} |p|^{-1}. 
\label{eq:b4}
\end{equation}
Since $(1+|x|)^{-2-\delta}$ is short--range, the operator $M$ is
compact. Denote by $\lambda_1, \lambda_2, \dots, \lambda_n, \dots$ 
and, respectively, $\varphi_1, \varphi_2, \dots, \varphi_n, \dots$ the
eigenvalues, respectively the eigenfunctions, of $M$. The eigenvalue
$\lambda$'s accumulate to zero. Thus, $\lambda_n > -1/4$ for all $n$
sufficiently large. Using the spectral decomposition of $M$ we can 
estimate 
\begin{equation}
I \equiv \frac{3}{4}(f,f) + (Mf,f) 
\ge \frac{3}{4} \vert| f \vert|^2 + \sum_{i=1}^{n-1} \lambda_i
|(f,\varphi_i)|^2 + \lambda_n \vert|f_{\perp} \vert|^2,
\label{eq:b5}
\end{equation}
for any $f \in L^2(\rz^3)$, with $f_{\perp}$ being the projection of $f$
onto the subspace orthogonal to $\varphi_i$,
all $1 \le i \le n-1$. 
Collecting terms, we get,
\begin{eqnarray}
I &\ge \left(\frac{3}{4} + \lambda_n \right) \vert|f_{\perp} \vert|^2
+
\sum_{i=1}^{n-1} \left(\frac{3}{4} + \lambda_i \right) \vert (f,\varphi_i)
\vert^2 \nonumber \\
& \ge \frac{1}{2} \vert| f_{\perp} \vert|^2 + \sum_{i=1}^n 
c_i \vert(f, \varphi_i) \vert^2, 
\label{eq:b6}
\end{eqnarray}
with $c_i=(3/4)+\lambda_i$. Now set $f= \vert P \vert \psi_1$. Since
$f$ is supported in the region $\vert P \vert < \alpha^q$, and the 
$\varphi_i$'s are independent of $\alpha$, 
$$
\frac{|(f,\varphi_i)|^2}{\vert| f \vert|^2} \to 0, \qquad
\mbox{and}
\qquad
\frac{\vert| f_{\perp} \vert|^2}{\vert| f \vert|^2} \to 1
$$
as $\alpha \to 0$. This finally implies that
\begin{equation}
L_1[\psi_1] \ge E_0 \vert| \psi_1 \vert|^2.
\label{eq:b7}
\end{equation}

\subsection{Intermediate momentum estimates}

Finally we have to estimate $L_2[\psi_2]$. 
We start with
\begin{equation}
L_2[\psi_2] \ge (T\psi_2,\psi_2) + (\lambda_0 - \frac{\gamma}{2})
(V \psi_2,\psi_2) - \kappa c_L((1+\vert x \vert)^{-2-\delta} \psi_2,\psi_2)
\label{eq:c1}
\end{equation}
with $\kappa < \gamma/16$.

Using estimates similar to (\ref{eq:a3.1},\ref{eq:a3.2}), 
we get for an arbitrary $\epsilon >0$,
and some $C_0(\epsilon)$ (provided $\alpha$ is sufficiently small),
\begin{equation}
(T \psi_2, \psi_2) \ge - (1-\epsilon) (\Delta_x \psi_2, \psi_2) + 
(1-\epsilon) (H_f, \psi_2, \psi_2) - C_0(\epsilon) \alpha
\vert| \psi_2 \vert|^2 + E_0 \vert| \psi_2 \vert|^2,
\label{eq:c2}
\end{equation}
(here we used the fact that $E_0 \le C \alpha$ for small $\alpha$ and 
some constant $C$ independent of $\alpha$). 

We first assume
\begin{equation}
(H_f \psi_2, \psi_2) \ge C_0(\epsilon) \alpha \vert| \psi_2 \vert|^2 
\frac{1}{1-\epsilon}.
\label{eq:c3}
\end{equation}
In this case
\begin{eqnarray}
& L_2[\psi_2] \ge E_0 \vert| \psi_2 \vert|^2 - (1-\epsilon) 
(\Delta_x \psi_2,\psi_2) + (\lambda_0 - \frac{\gamma}{2}) (V \psi_2,\psi_2) 
\nonumber
\\& - \kappa c_L((1+\vert x \vert)^{-2-\delta} \psi_2,\psi_2) 
\ge E_0 \vert| \psi_2 \vert|^2,
\label{eq:c4}
\end{eqnarray}
provided $\epsilon < \gamma/(4 \lambda_0)$. 

To complete the proof of the theorem it suffices to consider the case,
\begin{equation}
(H_f \psi_2, \psi_2) \le C_0(\epsilon) \alpha \vert| \psi_2 \vert|^2 
\frac{1}{1-\epsilon}.
\label{eq:c5}
\end{equation}
Notice that $(H_f \psi_2, \psi_2) \ge (|P_f| \psi_2, \psi_2)$ and
$$
\nabla_x^2 = p_e^2 = (P-P_f)^2 = |P|^2 - 2 P P_f + |P_f|^2 \ge |P|^2 - 2
|P| |P_f|.
$$
On the support of $\psi_2$, $\alpha^q \le |P| \le P_0$. Thus
\begin{eqnarray}
& -(1-\epsilon) (\Delta_x \psi_2, \psi_2) - 
C_0(\epsilon)\alpha \vert| \psi_2 \vert|^2 \ge \nonumber \\
& \ge 
(1-\epsilon) (|P|^2 \psi_2,\psi_2) -2(1-\epsilon)(|P||P_f| \psi_2,\psi_2)
-C_0(\epsilon)\alpha \vert| \psi_2 \vert|^2 \ge 
\nonumber \\
& \ge
(1-\epsilon) (|P|^2 \psi_2,\psi_2) -2(1-\epsilon) P_0 (H_f \psi_2,\psi_2)
-C_0(\epsilon)\alpha \vert| \psi_2 \vert|^2 \ge \nonumber \\
& \ge (1-\epsilon) (|P|^2\psi_2,\psi_2) - C \alpha \vert| \psi_2 \vert|^2,
\end{eqnarray}
with some constant $C$ independent of $\alpha$.
Combining the last estimate with (\ref{eq:c4}) 
we arrive at
\begin{eqnarray}
& L_2 [\psi_2] \ge (1-\epsilon)(|P|^2 \psi_2, \psi_2) - C \alpha 
\vert| \psi_2 \vert|^2 + E_0 \vert| \psi_2 \vert|^2 \nonumber \\
& + (\lambda_0 - \frac{\gamma}{2}) (V \psi_2, \psi_2) - \kappa c_L
((1+|x|)^{-2-\delta} \psi_2, \psi_2).
\label{eq:c7}
\end{eqnarray}

Proceeding as in the estimates for large momenta, we introduce 
relative coordinates 
for photons. Let $\tilde \psi_2$ be $\psi_2$ written in these coordinates. 
Then, 
\begin{eqnarray}
& L_2 [\psi_2] = (1-\epsilon) (|P|^2 \tilde \psi_2, \tilde \psi_2)
- C \alpha \vert|\tilde \psi_2 \vert|^2 + E_0 \vert| \psi_2 \vert|^2 + 
\nonumber
\\
& (\lambda_0 - \frac{\gamma}{2}) (V \tilde \psi_2, \tilde \psi_2) - 
\kappa c_L ((1+|x|)^{-2-\delta} \tilde \psi_2, \tilde \psi_2).
\end{eqnarray}
Recall that
$$
-\frac{\gamma}{4 \lambda_0} (\vert P \vert^2 \tilde \psi_2, \tilde \psi_2)
- \kappa c_L ((1+|x|)^{-2-\delta} \tilde \psi_2, \tilde \psi_2) \ge 0,
$$
and
\begin{eqnarray}
& L_2[\psi_2] \ge (1-\epsilon - \frac{\gamma}{4 \lambda_0})
(|P|^2 \tilde \psi_2, \tilde \psi_2) - C \alpha \vert| \tilde \psi_2 \vert|^2 +
\nonumber \\
& + E_0 \vert| \psi_2 \vert|^2 + (\lambda_0 - \frac{\gamma}{2})
(V \tilde \psi_2, \tilde \psi_2).
\end{eqnarray}
Due to the condition $\vert P \vert > \alpha^q$, with $q<1/2$, for small 
$\alpha$ we have
$$
\frac{\gamma}{8 \lambda_0} 
(|P|^2 \tilde \psi_2, \tilde \psi_2) \ge \frac{\gamma}{8 \lambda_0}
\alpha^{2q} \vert| \tilde \psi_2 \vert|^2 > C \alpha 
\vert| \tilde \psi_2 \vert|^2.
$$

To complete the proof of the theorem  it suffices now to notice that
$$
(1-\epsilon - \frac{3}{8 \lambda_0} \gamma) 
(|P|^2 \tilde \psi_2, \tilde \psi_2) +(\lambda_0 - \frac{\gamma}{2}) 
(V\tilde \psi_2, \tilde \psi_2) \ge 0,
$$
for $\epsilon < \gamma/(8 \lambda_0)$. 

\section{Acknowledgements}
One of us (S. Vugalter) thanks the hospitality of the Physics
Department, P. U. Cat\'olica de Chile, where part of this 
work was done.

\bibliographystyle{amsalpha}

\end{document}